\documentstyle[prb,aps,epsfig,twocolumn]{revtex}
\begin{document}
\draft
\flushbottom
\twocolumn[
\hsize\textwidth\columnwidth\hsize\csname @twocolumnfalse\endcsname

\title{Trace Formulas and Bogomolny's Transfer Operator}
\author{Oleg Zaitsev, R. Narevich, and R. E. Prange}
\address{Department of Physics, University of Maryland, College Park,
Maryland 20742}
\maketitle

\tightenlines
\widetext
\advance\leftskip by 57pt
\advance\rightskip by 57pt
\begin{abstract}
The trace formulas commonly used in discussing spectral properties of
quantum or wave systems are derived simply and directly from the Bogomolny
transfer operator. Special cases are the Gutzwiller formula, the Berry-Tabor
formula, and the perturbed Berry-Tabor formula. A comment is made about
interpolation formulas proposed in the latter case.
\end{abstract}


]

\narrowtext
\tightenlines

\section{Introduction}

The Gutzwiller trace formula \cite{Gutz} and its relatives are the
cornerstone of much of the subject of `quantum chaology'. This formula gives
a formal quasiclassical expression for the density of states, in other
words, for the energy levels, of a `hard' chaotic system. The Berry-Tabor
trace formula \cite{BT1,BT2} is the corresponding formula for an integrable
system. These are descendants of the exact Selberg\cite{Selberg} trace
formula, which applies to spaces of negative curvature.

Twenty-five years of work on the trace formulas leads to the conclusion that
they are not especially useful for finding actual energy levels. This is
because they are divergent formulas unless they are smoothed over energy. On
the other hand, many interesting observables can be expressed in terms of
the density of states smoothed over some energy scale. For example, the
specific heat or magnetic susceptibility are observables of this type. The
smoothing can come about because of finite temperature or finite
experimental resolution, say. With appropriate smoothing, the trace formulas
are very useful.

We here provide a derivation of these and related formulas based on
Bogomolny's transfer operator \cite{Bog}. One of the main virtues of the
transfer operator is that it can be used to study energy levels,
wavefunctions and scattering amplitudes. The possibility of obtaining the
trace formulas from the transfer operator is widely known, and certainly was
to Bogomolny, especially in the Gutzwiller case. 

An explicit derivation of the Berry-Tabor formula or of the perturbed
Berry-Tabor formula that is based on the Bogomolny operator does not
seem to have appeared in print, however. That is the main purpose of
this paper. We restrict consideration to the two-dimensional case, which
is by far the most important in practice, and to a closed system with a
discrete spectrum.

The original derivation of the Gutzwiller formula\cite{Gutz} was based
on a stationary phase evaluation of Feynman's path integral expression
for the quantum propagator. This reduces the propagator from a sum over
all paths to a sum over just the classical paths. Although the Feynman
formulation gives a persuasive picture of the relationship of quantum to
classical mechanics, it does have some defects. For one thing, it is
mathematically not very well defined. Also, its perspective on classical
mechanics is very elementary. In particular, it does not incorporate the
insights gained by use of Poincar\'e's surface of section method. The
present derivation removes these defects.

The Berry-Tabor formula was derived in several ways \cite{Gutz2,BT1,BT2}. For
example, the density of states expressed as a sum over explicitly known energy
levels can be transformed by use of the Poisson sum formula. Action-angle
variables are also useful to obtain this formula \cite{BT1}.

With considerable generality, for a given problem, that is, for a given
Hamiltonian, there exists \cite{DorSmiDie} a mathematically well defined
kernel or operator $K(q,q^{\prime },E)$, whose arguments $q,q^{\prime }$ are
on a one-dimensional {\em Poincar\'e surface of section } [SS]. The exact
spectrum is determined by the Fredholm integral equation 
\begin{equation}
\psi (q)=\int_{SS}dq^{\prime }\,K(q,q^{\prime },E)\psi (q^{\prime }),
\label{Fred1}
\end{equation}
which has nontrivial solutions only if $E$ is on the spectrum. The integral
runs over the surface of section. This is an exact quantum version of the
classical SS method. A well known and much used example\cite{GeoPran} is a
billiard where the surface of section is the boundary and $K(q,q^{\prime
},E) $ is a Hankel function with argument $kL(q,q^{\prime }).$ Here $k=\sqrt{%
2mE}/\hbar $ with $m$ the mass and $E$ the energy of the particle in the
billiard, and $L$ is the chord distance between points on the boundary
specified by the $q$'s. The exact or numerically approximated method is in
this case called the {\em boundary integral method} \cite{Bog}.

\section{Bogomolny's transfer operator, $T$}

In more general cases, $K$ does not have a simple expression. However,
Bogomolny\cite{Bog} in effect argued that in quasiclassical approximation $K$
may be replaced by its asymptotic form for large separation of points $
q,q^{\prime }$. In this approximation, the operator, called $T,$ is quite
simple, because it involves only a few, often just one, short orbits.
Namely, 
\begin{eqnarray}
T(q,q^{\prime },E) &=&\left( \frac 1{2\pi i\hbar }\left| \frac{\partial
^2S(q,q^{\prime },E)}{\partial q\partial q^{\prime }}\right| \right) ^{\frac 
12}  \nonumber  \label{T} \\
&&\ \times \exp \left[ \frac i\hbar S(q,q^{\prime },E)+\frac{i\pi }2\nu
\right] ,  \label{T}
\end{eqnarray}
where $S$ is the action $\int pdq$ along the classical path of energy $E$,
which leaves the surface of section at $q^{\prime }$ and returns to it {\em %
for the first time} at point $q.$

Care must be taken with the phase attributed to the prefactor and with
phase changes encountered at points along the path where the leading order
semiclassical approximation fails, e.g. at a billiard boundary. These are
incorporated into the `Maslov' phase $\nu .$ It can be regarded as part of
the action $S$, i.e. $S\rightarrow S+h\nu /4$ and thus is a quantum
correction. Also, in doing integrals such as those below, contributions to
$ \nu $ may be obtained. We shall, however, suppress $\nu $ in the sequel,
since that is not the point of this work.

It should be noted that $S$ is the {\em generator of the classical surface
of section map}, namely $p=\partial S(q,q^{\prime })/\partial q, $ $
p^{\prime }=-\partial S/\partial q^{\prime }$, where $p, $ $ p^{\prime }$
are the momenta of the orbit parallel to the surface of section at
$q,q^{\prime }$. $ S$ thus contains {\em all the long-time and long-orbit
information }of the classical mechanics.

\section{The general quasiclassical trace formula}

Eq. (\ref{Fred1}), with $K$ replaced by $T$, has a solution when the
Fredholm determinant \cite{Smith}
\begin{equation}
D(E)=\det \left[ 1-T(E)\right] =0.  \label{FredD}
\end{equation}
The density of states is given by the logarithmic derivative \cite
{Bog,GeoPran} 
\begin{equation}
d(E)-\bar d(E)\equiv d_{osc}(E)=\frac{-1}\pi 
\mathop{\rm Im}
\left[ \frac{d\ln D(E+i\epsilon )}{dE}\right] ,  \label{dosc}
\end{equation}
where $d(E)=\sum_a\delta (E-E_a)$ and $\bar d(E)$ is the smoothed (Weyl)
density of states. Using the relationship $\ln \det (1-T)=%
\mathop{\rm Tr}
\ln (1-T),$ and expanding the logarithm as if $T$ were small we have 
\begin{equation}
d_{osc}(E)=\frac 1\pi 
\mathop{\rm Im}
\sum_{n=1}^\infty \frac 1n\frac{d\tau _n(E)}{dE},  \label{dexp}
\end{equation}
where 
\begin{equation}
\tau _n(E)=%
\mathop{\rm Tr}
T(E)^n.  \label{taun}
\end{equation}

Since $T$ is in a quasiclassical sense close to a unitary operator, some
of the eigenvalues of $T$ are close to the unit circle, which makes the
expansion (\ref{dexp}) divergent for real $E$. However, in this paper we
ignore this problem, since, as we said, in practice the trace formulas are
useful only when averages are taken, which suppress the contribution of
the terms with large $n.$

The trace formulas are obtained by evaluating the traces defining the
$\tau _n$'s in stationary phase approximation, if this is meaningful.
There are $n$ integrals to be carried out, i.e.  
\begin{equation} 
\tau _n=\int \ldots \int dq_1\ldots dq_nT(q_1,q_2)\ldots T(q_n,q_1).
\label{tauint}
\end{equation} 
The fundamental composition relation
\begin{equation} 
T_2(q,q^{\prime })=\int d\bar q\,T(q,\bar q)T(\bar
q,q^{\prime }) \label{tau2} 
\end{equation} 
evaluated in stationary phase is given by \cite{Bog}
\begin{eqnarray} 
T_2(q,q^{\prime }) &=&\sum_p\left( \frac 1{2\pi i\hbar }\left| \frac{
\partial ^2S_{2p}(q,q^{\prime },E)}{\partial q\partial q^{\prime }}\right|
\right) ^{\frac 12} \nonumber \label{T2} \\ &&\ \times \exp \left[ \frac
i\hbar S_{2p}(q,q^{\prime },E)\right] , \label{T2}
\end{eqnarray} 
where 
\begin{equation}
S_{2p}(q,q^{\prime },E)=S(q,q_p)+S(q_p,q^{\prime })\text{ } \label{S2}
\end{equation} 
is the action of a classical orbit from $q^{\prime }$ to $q$ at energy
$E$, which arrives at the surface of section at $q$ for the {\em second}
time after having crossed the {\em first} time at $q_p.$ The stationary
phase condition is $\partial [S(q,q_p)+S(q_p,q^{\prime })]/\partial
q_p=0.$ The solution of this equation, $q_p$, is a function of $q$,
$q^{\prime }$. There are, in general, more than one solution $q_p$ to this
stationary phase condition, which can lead to an exponential proliferation
of orbits as a function of $n$. Thus, a $T$ operator corresponding to
multiple SS crossings can be defined, and it has exactly the same form as
Eq. (\ref{T}), although there are in general many terms of this type.

There are two common situations. In the first, all $n$ integrals may be
performed by stationary phase. This is the Gutzwiller assumption. In the
second case, the first $n-1\,$ integrals are well approximated by
stationary phase but this method fails for the last integral.

Eq. (\ref{T2}) may be iterated to obtain the final integral
\begin{eqnarray} 
\tau _n &=&\int dq\sum_p\left( \frac 1{2\pi i\hbar }\left| \frac{\partial
^2S_p(q,q^{\prime },E)}{\partial q\partial q^{\prime }}\right| \right)
_{q=q^{\prime }}^{\frac 12} \nonumber \label{taunf} \\
&&\ \times \exp \left[ \frac i\hbar S_p(q,q,E)\right] .  \label{tn}
\end{eqnarray} 
Here $p$ denotes a `closed' orbit from $q$, which crosses the SS $n-1$
times and arrives back at $q$ at its $n$th encounter with the SS.

\section{The Gutzwiller trace formula}

In the Gutzwiller case, the final $q$ integral may be done by stationary
phase also, that is, the phase $S_p(q,q,E)/\hbar $ is a rapidly varying
function of $q.$ This picks out as stationary points the solutions of $
\partial S_p(q,q^{\prime },E)/\partial q+\partial S_p(q,q^{\prime
},E)/\partial q^{\prime }=p-p^{\prime }=0$ at $q=q^{\prime }=q^{*}.$ Thus
periodic orbits are selected. Let $q,q^{\prime }$ be in the vicinity of
$q^{* \text{ }}$ and expand
\begin{eqnarray}
S_p(q,q^{\prime }) &\simeq &S_p(E)+p^{*}(\delta q-\delta q^{\prime }) 
\nonumber  \label{Sp} \\
&&+\frac 12W_{11}\delta q^2+W_{12}\delta q\delta q^{\prime }+\frac 12%
W_{22}\delta q^{\prime 2},  \label{Sp}
\end{eqnarray}
where $\delta q=q-q^{*}$, $\delta q^{\prime }=q^{\prime }-q^{*}$ and
$p^{*}$ is the momentum of the periodic orbit. The $W$ matrix depends on
energy and on the orbit. Let the orbit $p$ of `length' $n$ (i.e. returning
to the surface of section $n$ times) consist of $r$ repetitions of
`length' $s, $ with $n=rs. $ The $q$ integral is then done to obtain the
contribution of this periodic orbit to the trace
\begin{equation}
\tau _p=\ s\left| \frac{W_{12}}{W_{11}+2W_{12}+W_{22}}\right| ^{\frac 12%
}\exp \left[ \frac i\hbar S_p(E)\right] .  \label{taurp}
\end{equation}
The factor $s$ arises because the stationary point for the last integral
can occur (or, in other words, the starting/ending point for the periodic
orbit can be chosen) at any of $s$ surface of section crossings.

The prefactor is usually expressed in terms of the monodromy matrix $M_p$
of orbit $p$, which connects the final momentum-position $p-p^{*}=\delta
p=W_{11}\delta q+W_{12}\delta q^{\prime }$ and $\delta q$ to the initial $
\delta p^{\prime }=-W_{12}\delta q-W_{22}\delta q^{\prime }$ and $\delta
q^{\prime }.$ That is
\begin{equation}
M_p=\left( 
\begin{tabular}{cc}
$-\frac{W_{11}}{W_{12}}$ & $\frac{W_{12}^2-W_{11}W_{22}}{W_{12}}$ \\ 
$-\frac 1{W_{12}}$ & $-
\frac{W_{22}}{W_{12}}
$
\end{tabular}
\right),  \label{Mp}
\end{equation}
so that the prefactor in Eq. (\ref{taurp}) can be written $\left| \det
(M_p-1)\right| ^{-1/2}.$

In the case of a repeated orbit, $S_p=rS_s,$ and $M_p=M_s^r.$ In taking
the derivative of $\tau _p$ only the rapidly varying phase needs to be
differentiated, giving a factor of $ dS_s(E)/dE=T_s,$ the period of the
primitive orbit, leading to the final expression
\begin{equation}
d_{osc}=\sum_{r,s}\frac{T_s/\hbar }{\left| \det (M_s^r-1)\right| ^{1/2}}\cos
\left\{ r\left[ \frac{S_s(E)}\hbar +\frac \pi 2\nu _s\right] \right\} ,
\label{doscG}
\end{equation}
in which we restore the Maslov index.

\section{The Berry-Tabor formula}

The above formula does not apply to integrable systems. In that case, the
periodic orbits are not isolated. The eigenvalues $\lambda $ of the
monodromy matrix are, for isolated orbits, $\lambda =e^{\pm \gamma },$
where for an unstable orbit the Lyapunov exponent $\gamma >0,$ and for a
stable orbit $\gamma $ is pure imaginary. For an integrable system, $
\gamma$ vanishes and formula (\ref{doscG}) is infinite.

In the case of an integrable system, it is easiest to go to action-angle
variables \cite{BT1}, $\theta ,I,\Theta ,J$ using as surface of section $
\Theta = 0, $ and energy $E=H(I,J)$ fixed. Then the action of the $T$
operator becomes $ S(q,q^{\prime })\rightarrow S(\theta -\theta ^{\prime
}).$ The replacement of the action depending on two variables separately
by a function of the difference of coordinates only characterizes the
integrable system. In fact, the action is
\begin{equation}
S(\theta -\theta ^{\prime },E)=(\theta -\theta ^{\prime })I+2\pi J.
\label{SInt}
\end{equation}
The actions $I,J$ specify the invariant torus, on which the orbit remains.
The frequencies of rotation around this torus are $\omega_I = \partial H /
\partial I$ and $\omega_J = \partial H / \partial J$. We may equally well
use energy $E$ and winding number $\omega\equiv \omega _I/\omega
_J=(\theta -\theta ^{\prime })/2\pi $ as variables to specify the orbit.
[This fails for harmonic oscillators where the frequencies are constants.
Modifications are necessary in that case \cite{BT1}.] Thus, $I,J$ are
regarded as functions of $E,\theta -\theta ^{\prime }$. It is easy to see
that $\partial S/\partial \theta =I.$ Obviously, the eigenfunction of the
$T$ operator is in this case $\psi =\exp iI\theta .$

The first $n-1$ integrals may be carried out by stationary phase, as
before, giving a result $S_p(\theta -\theta ^{\prime })\,$ where $\theta
,$ which is physically equivalent to $\theta ^{\prime }$, is set equal to
$\theta ^{\prime }+2\pi m.$ The winding number is $m/n$, i.e. it is
rational. Clearly $S_p(\theta -\theta ^{\prime })=nS\left( \Delta \theta
\right)$, where $ \Delta \theta =(\theta -\theta ^{\prime })/n.$ Remark
first that in this limit $p$ is a periodic orbit. It is a periodic orbit
for each $\theta $ with identical period and other properties, in other
words, this describes a set of periodic orbits, which are not isolated.

The final integral Eq. (\ref{tn}) is then trivial, giving a factor $2\pi
.$ Note a factor $ i^{-\frac 12}$, which remains and gives a shift $\pi
/4$ in the final answer below. The prefactor involves the second
derivative (at constant $E$), $ S_p^{\prime \prime }=\frac 1n\partial
I/\partial \Delta \theta $. This has traditionally been expressed in terms
of the function $g_E(I)=J$, which determines the action $J$ given the
energy and $I$. One finds that $ g_E^{\prime }=-\Delta \theta /2\pi,$ the
derivative being taken at fixed $ E$. Therefore $S_p^{\prime \prime
}=-(2\pi ng_E^{\prime \prime })^{-1}.$ Putting this all together we obtain
the standard result \cite{BT1,BT2}
\begin{equation}
d_{osc}=\sum_p\frac{T_p}{\pi \hbar ^{3/2}n^{3/2}\left| g_E^{\prime \prime
}\right| ^{1/2}}\cos \left( \frac{S_p}\hbar +\frac \pi 2\nu _p-\frac \pi 4
\right).  \label{dosc}
\end{equation}

There are other cases, in which the last integral Eq. (\ref{tn}) cannot be
done by stationary phase. There may be particular sets of nonisolated
orbits in an otherwise chaotic system. Prominent examples are the bouncing
ball orbits in the stadium \cite{Stad} or Sinai billiard \cite{Sinai}.
There are also nonisolated orbits in the ray splitting billiard
\cite{RaySplit}.

\section{Perturbed Berry-Tabor formula}

Another important example is the perturbed integrable system. Then the
action defining the $T$ operator has the form $S(\theta -\theta ^{\prime
})+\epsilon W_1(\theta ,\theta ^{\prime }),$ and $\epsilon <<1.$ To the
leading order in $\epsilon $, we may assume that the stationary phase
points of the first $n-1$ integrals are not shifted. The last integral
then involves
\begin{equation}
I_W=\frac 1{2\pi }\int d\theta \exp \left[ \frac{i\epsilon }\hbar \hat W
(\theta )\right] ,  \label{IW}
\end{equation}
where $\hat W(\theta )= \sum_{r=1}^n W_1(\theta - \theta_{r-1},\theta -
\theta_r)$, and $\theta_r = 2 \pi m r/n$ for the winding number $m/n$. It
can be shown that $\hat W(\theta) = W_n (\theta, \theta - 2 \pi m)$, where
$W_n (\theta, \theta^{\prime})$ is the first order correction to the
action for the orbit that returns to the SS for the $n$th time.  The
perturbed result is just the Berry-Tabor result with the substitution
$\cos \phi \rightarrow
\mathop{\rm Re}
\left[ I_W\exp (i\phi )\right] $, where $\phi $ is the argument of cosine
in Eq. (\ref{dosc}) \cite{OdA}. 

Clearly, $I_W$ is a sort of generalized Bessel function. Formulas for
$I_W$ interpolating between $\epsilon =0$ and $\epsilon /\hbar $ large
have been given in the literature. The first were based on the idea that
often $\hat W$ will be well approximated by the first terms of its Fourier
series \cite{OdA}, e.g. $\hat W\approx w_0+w_1\cos (\theta -\theta _0).$
The phase shift $ \theta _0$ is of no importance, and $I_W\approx \exp
(i\epsilon w_0/\hbar )J_0(\epsilon w_1/\hbar ),$ where $J_0$ is the Bessel
function.

Interpolation formulas along this line using more parameters have been given
in the literature\cite{Ullmo}. We give an alternative four parameter formula
in the Appendix, since we believe the existing formulas are defective.

One comment concerning these formulas based on a small parameter $\epsilon
$ is in order. Both {\em classical} perturbation theory and {\em quantum}
results based on small $\epsilon $ have relatively large effects,
proportional to $\sqrt{\epsilon }$, when the system is in the neighborhood
of a resonance. Furthermore, the trace formulas apply precisely to the
resonant cases.

Nevertheless, it is the case that for the perturbed trace formula, the
corrections incorporated into Eq. (\ref{IW}) can be expanded in powers of $
\epsilon ,$ and such a series is convergent.

The reason is as follows: To find energy levels and wave functions, one
must solve Eqs. (\ref{FredD},\ref{Fred1}). This, in effect, depends on
the very high terms and convergence properties of the trace formulas.
The $\sqrt{\epsilon } $ appears if one is interested in this level of
accuracy \cite{PNZ}. However, to look at long range energy level
correlations, and the smoothed density of states, it is only necessary
to consider the leading terms in the trace formula expansion. These
leading terms can be calculated from short periodic orbits. Already
classically, it is known that short time quantities can be calculated
perturbatively as a power series in $\epsilon $.

\section{Conclusion}

The Bogomolny transfer operator $T$ provides a unified approach to the
quasiclassical description of wave systems, without needing to make
specific reference to the chaotic or integrable properties of the
system. Its greatest importance is, no doubt, that it provides a method
to find actual energy levels and wave functions. However, it can also be
used to find the trace formulas, which are widely used in this field.
This derivation of the trace formulas is quite easy and direct and it is
on at least as firm a mathematical footing as most other derivations.

\appendix 

\section{Uniform approximation}

Suppose $\hat W(\theta )$ has a single maximum and minimum at $\theta
=\theta _x,\theta _n$, respectively. For large $\epsilon /\hbar $ the
integral (\ref{IW}) can be done by stationary phase. This is sometimes
called the Gutzwiller case because the continuum of identical periodic
orbits has turned into two isolated orbits. The integral in this
approximation depends on four parameters, namely, the magnitudes of $\hat
W$ and $\hat W^{\prime \prime }$ at the extrema. It does not depend
directly on the position of the extrema. Thus, in order to have the
correct large $ \epsilon /\hbar $ limit, one should use a four parameter
expression for $I_W$.

Inspired by Ref. \cite{Ullmo}, [UGT], we parametrize $\hat W(\theta
)=W^{(0)}+W^{(1)}\cos [\xi (\theta )]$, where $\xi (\theta _x)=0$ and $\xi
(\theta _n)=\pi .$ Note that $W^{(0)}$ and $W^{(1)}$ are not necessarily
equal to $w_{0, \text{ }}w_1$. We may as well take $\theta _x=0$ and
$\theta _n=\pi $, to simplify things. Now take $\xi $ as the solution of
$\theta = \xi -A\sin \xi -B\sin 2\xi .$ This goes beyond UGT, who take
$B=0.$ The derivative is $ \hat W^{\prime }=-\xi ^{\prime }(\theta
)W^{(1)}\sin \xi$, and the second derivative, evaluated at an extremum,
is $\hat W^{\prime \prime }=-(\xi ^{\prime })^2 W^{(1)}\cos \xi.$ Since
$d\theta /d\xi =1-A\cos \xi -2B\cos 2\xi ,$ we find $\xi ^{\prime
}(0)=\left[ 1-A-2B\right] ^{-1}$ and $\xi ^{\prime }(\pi )=\left[
1+A-2B\right] ^{-1}.$ We may solve for $A$, $B$, $W^{(0)},$ and $ W^{(1)}$
to obtain a parametrized version of $\hat W$, which interpolates between
$\epsilon /\hbar =0$ and $\epsilon /\hbar $ large.

The integral Eq. (\ref{IW}) now is 
\begin{eqnarray}
I_W &=&\frac{I_0}{2\pi }\int d\xi (1-A\cos \xi -2B\cos 2\xi )  \nonumber
\label{IWA} \\
&&\times \exp \left[ i\left( \epsilon /\hbar \right) W^{(1)}\cos \xi
\right] ,
\label{IWA}
\end{eqnarray}
where $I_0=\exp \left[ i\left( \epsilon /\hbar \right) W^{(0)}\right] .$
The integrals may be expressed as Bessel functions $J_m$, whose arguments
are $\epsilon W^{(1)}/\hbar$, giving
\begin{equation}
I_W=I_0(J_0-iAJ_1+2BJ_2).  \label{IWF}
\end{equation}

As an exercise, and to compare with UGT, we calculate 
$I_{W\text{ }}$ for coupled quartic oscillators. The Hamiltonian for this
case is 
\begin{equation}
H=\frac 12(p_x^2+p_y^2)+ax^4+by^4+\epsilon x^2y^2.  \label{Ham}
\end{equation}
The correspondence with UGT goes as follows: $a\leftrightarrow a(\lambda
)/b$, $b\leftrightarrow a(\lambda )b$, $\epsilon \leftrightarrow 2\lambda
a(\lambda )$, where $a(\lambda )$ is a specified function of $\lambda .$
However, we may choose the unit of length such that $a(\lambda )=1,$ i.e.
$ a=1/b.$ Thus $\lambda ,$ the small parameter of UGP, is in effect our $
\epsilon /2.$

Using action-angle variables, with 
$\Delta \Theta =2\pi $, we find 
\begin{equation}
S(\theta -\theta ^{\prime })=\frac{4KE^{3/4}}{3\pi }\left[ \frac{(\theta
-\theta ^{\prime })^4}a+\frac{(2\pi )^4}b\right] ^{1/4},  \label{Squart}
\end{equation}
where $K\equiv K(m=1/2)$ is the complete elliptic integral of the first
kind \cite{Abram}. Here $E$ is the energy of the unperturbed system. The
perturbing action $W_n$ is found (to leading order in $\epsilon )$ in
terms of Jacobi elliptic functions \cite{Abram} as
\begin{eqnarray}
W_n(\theta ,\theta ^{\prime }) &=&-\frac{KE^{3/4}}{4\pi (a b)^{1/4} }\frac
\omega {\left[ a+b\omega ^4\right] ^{3/4}} \label{Wquart} \\
&&\ \times \int_{-2\pi n \omega }^0d\theta ^{\prime \prime
}\text{sd}^2[\alpha (\theta ^{\prime \prime }+\theta )]\space
\text{sd}^2\left[ \alpha \theta ^{\prime \prime }/\omega \right] ,
\nonumber
\end{eqnarray}
where $\alpha =2K/\pi $ and $\omega =\left( \theta -\theta ^{\prime
}\right) /2\pi n.$ We focus on the 1-1 resonance, $\omega =1$, and $\hat
W(\theta )=W_1(\theta ,\theta -2\pi )$. Note that the shape of $\hat
W(\theta )$ for fixed $\omega $ is independent of $a$ and $E$ and depends
only on the winding number.

\begin{figure}[tbp]
{\hspace*{0.7cm}\psfig{figure=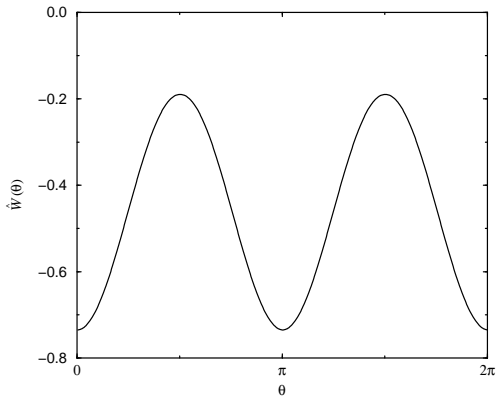,height=6.5cm,width=5.35cm,angle=270}}
{\vspace*{.13in}}
\caption[ty]{ The function $\hat W(\theta )$ vs. $\theta $ for
anharmonic oscillators with $x^2y^2$ coupling, for the (1,1) resonance.
The parameters $a,$ $b,$ and $E$ are defined in the text and each is
taken to be unity.}
\label{fig:1}
\end{figure}

\begin{figure}[tbp]
{\hspace*{0.7cm}\psfig{figure=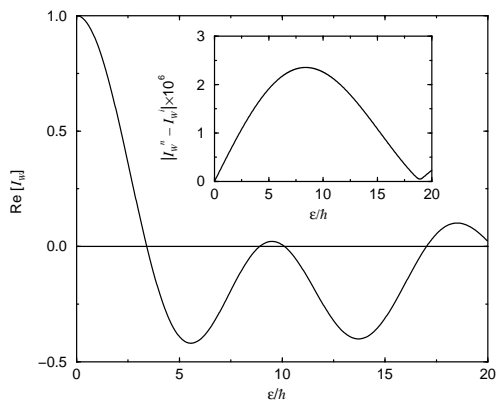,height=6.5cm,width=5.35cm,angle=270}}
{\vspace*{.13in}}
\caption[ty]{The real part of $I_W$ as a function of $\epsilon /\hbar .$
The difference between this function computed by numerical quadrature
$(I_W^n)$ and from the interpolation formula $(I_W^i)$ is shown in the
inset.}
\label{fig:2}
\end{figure}

\begin{figure}[tbp]
{\hspace*{0.7cm}\psfig{figure=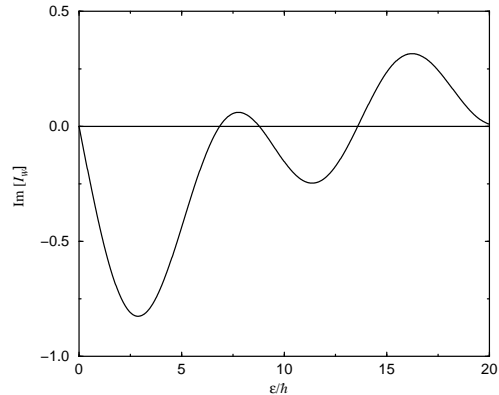,height=6.5cm,width=5.35cm,angle=270}}
{\vspace*{.13in}}
\caption[ty]{The imaginary part of $I_W$ as a function of $\epsilon /\hbar
.$}
\label{fig:3}
\end{figure}

Fig. \ref{fig:1} gives a graph of $\hat W(\theta ).$ It has two maxima and
minima, which are symmetrically related. It is easy to extend the
interpolation expression to take this into account. $\text{Re}[I_W]$ as a
function of $ \epsilon /\hbar$ is presented in Fig. \ref{fig:2}, and
$\text{Im}[I_W]$ is shown in Fig. \ref{fig:3}. We find that the
interpolation formula and the exact integral agree to within a part in
$10^{-5}$ (inset Fig. \ref{fig:2}).

UGT assume $B$ is zero. This means that they cannot generally satisfy the
asymptotic condition for large $\epsilon /\hbar $. It can be seen that
$B=0$ implies a relation between the curvatures at minimum and maximum,
namely,
\begin{equation}
\sqrt{W^{(1)}/\hat W_{\min}^{\prime \prime }}+\sqrt{W^{(1)}/\left| \hat
W_{\max}^{\prime \prime }\right| }=2/r, \label{curvcond}
\end{equation}
where $r$ is the number of maxima (or minima) of $\hat W(\theta )$ per
$2\pi $ ($r=2$ for 1-1 resonance). However, for quartic oscillators
with $x^2y^2$ coupling, this is well satisfied, the {\em l.h.s.} of Eq.
(\ref{curvcond}) being equal to 1.00021. This probably accounts for the
good numerical results of UGT in this case. We find $A=-1.49\times
10^{-2}$, $B=-1.04\times 10^{-4}$, independent of the other parameters,
and $W^{(0)}=-0.46$ and $W^{(1)}=0.27 $ for $E=a=1/b=1.$

To obtain a fourth parameter, UGT formally take $g_E^{\prime \prime }$ as
an empirical parameter depending on $\epsilon ,$ arguing that an
independent evaluation of $g_E$ is rather laborious and time consuming.
However, $g_{E\text{ }}$ is a geometrical quantity depending on the
invariant tori, defined only at $\epsilon =0.$ Its definition cannot be
easily extended for $\epsilon >0,$ since the topology of the invariant
tori changes under perturbation. We found the simple formula, in the
present exercise,
\begin{equation}
g_E(I) = J = \left[ \left( \frac {4K} {3 \pi} \right)^{4/3} E - \left( \frac 
a b \right)^{1/3} I^{4/3} \right]^{3/4}.
\end{equation}
UGT do not mention the curvatures $\hat W_{\min}^{\prime \prime }$, $\hat
W_{\max}^{\prime \prime }$ explicitly. These parameters are replaced by the
determinants of the monodromy matrix $M_p - 1$, at the stable and unstable
orbits [see our Eq. (\ref{doscG})]. These determinants can be expressed in
terms of $\hat W^{\prime \prime }$ and $g_E^{\prime \prime }.$ Eliminating
thus $\hat W^{\prime \prime }$, we obtain the result of UGT in our notation,
\begin{eqnarray}
\det (M_u-1) &=&-c\epsilon g_E^{\prime \prime }W^{(1)}/[1-A+2B]^2,
\nonumber
\label{detM} \\
\det (M_s-1) &=&c\epsilon g_E^{\prime \prime }W^{(1)}/[1+A+2B]^2.
\label{detM}
\end{eqnarray}
Here $c$ is a known constant.

Given $g_E^{\prime \prime }$ and $W^{(1)}$, we can find $A$ and $B$ in
terms of the determinants. Setting $B=0$ leaves too few parameters, so
$g_E^{\prime \prime },$ extended to be a function of $\epsilon ,$ is
pressed into service by UGT. However, for small $\epsilon ,$ the
determinants are proportional to $\epsilon $ because the Lyapunov
exponents are proportional to $\sqrt{ \epsilon }.$ Their ratio is a
constant not necessarily equal to $-1$. The explicit $\epsilon $ cancels
out leaving $g_E^{\prime \prime }(\epsilon )$ as nearly constant for small
$\epsilon ,$ but not necessarily close to the zero $ \epsilon $ limit,
unless the system is such that $B$ happens to be small.

\end{document}